\documentclass[twocolumn,showpacs,preprintnumbers,amsmath,amssymb]{revtex4}
\usepackage{graphicx}
\usepackage{wasysym}
\usepackage{color}
\begin{document}
\title{A little inflation at the cosmological QCD phase transition}
\author{Tillmann Boeckel}
\affiliation{Institut f\"ur Theoretische Physik, Universit\"at Heidelberg\\ 
Philosphenweg 16, D-69120 Heidelberg, Germany}
\author{J\"urgen Schaffner-Bielich}
\affiliation{Institut f\"ur Theoretische Physik, Universit\"at Heidelberg\\ 
Philosphenweg 16, D-69120 Heidelberg, Germany}
\date{\today}

\begin{abstract} 
 We reexamine the recently proposed "little inflation" scenario that allows for a strong first order phase-transition of QCD at non-negligible baryon number in the early universe and its possible observable consequences. The scenario is based on the assumptions of a strong mechanism for baryogenesis and a quasistable QCD-medium state which triggers a short inflationary period of inflation diluting the baryon asymmetry to the value observed today. The cosmological implications are reexamined, namely effects on primordial density fluctuations up to dark matter mass scales of $M_{max} \sim 1 M_{\astrosun}$, change in the spectral slope up to $M_{max} \sim 10^6 M_{\astrosun}$, production of seeds for the present galactic and extragalactic magnetic fields and a gravitational wave spectrum with a peak frequency around $\nu_{peak} \sim 4 \cdot 10^{-8} \mbox{Hz}$. We discuss the issue of nucleation in more detail and employ a chiral effective model of QCD to study the impact on small scale structure formation.
\end{abstract}

\maketitle
\section{introduction}
\begin{figure}[ht]
	\centering
		\includegraphics[width=0.48\textwidth]{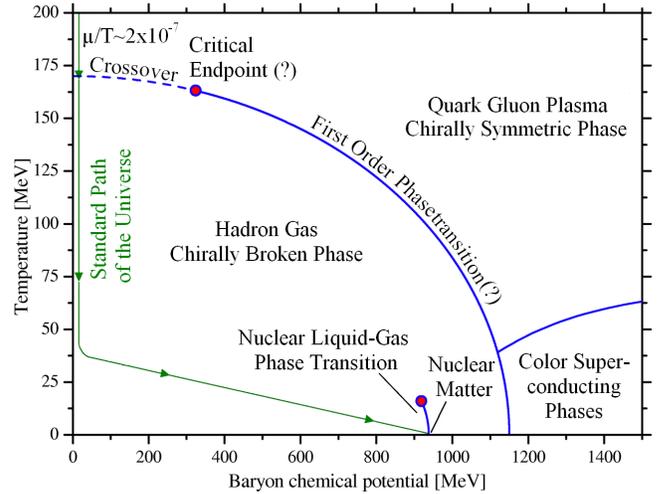}
		\caption{Sketch of a possible QCD-phase diagram with the commonly accepted standard evolution path of the universe as calculated e.g. in \cite{Fromerth02} depicted by the green path.}
	\label{fig:qcdpdstandard}
\end{figure}

\noindent At about 10 microseconds after the big bang a phase transition from the quark-gluon plasma to a hadron gas is expected to have taken place at a temperature of about  $T_{QCD} \approx 150-200$ MeV. In the last decade it has become more and more clear that this transition was most probably only a rapid crossover as indicated by more and more refined lattice gauge theory calculations at zero baryon density \cite{Aoki06b,Karsch07}. In standard cosmology the baryon asymmetry is tiny $\eta_B = n_B / s  \sim 10^{-9}$, with $n_B$ being the net baryon density and $s$ the entropy density, as deduced from later stages in the evolution of the universe. Therefore a first order QCD phase transition seemed very unlikely given the conditions. Still, the QCD phase diagram is for most parts terra incognita. The chiral and the deconfinement transition do not necessarily coincide but there are some indications from effective models \cite{Braun11} and lattice QCD calculations that there is at least a significant connection between the two. There has been recent progress in the attempt to include a finite baryon density on the lattice \cite{Fodor04,Karsch02} but effective models are still the method of choice to explore the uncharted regions of the QCD phase diagram \cite{PisarskiWilczek84}. Findings indicate that at finite baryon densities a first order phase transition can be expected as shown by chiral effective models of QCD \cite{StephanovRajagopalShuryak98} caused by the melting of quark and/or gluon condensates or by color superconductivity \cite{Alford08}. A sketch of a possible QCD phase diagram is depicted in figure \ref{fig:qcdpdstandard} along with the commonly accepted path the universe took during and after the QCD-transition. The universe starts out in the upper left and move along the temperature axis from the chirally symmetric quark gluon plasma through a crossover transition to the chirally broken hadron gas phase. Once protons and anti-protons stop to annihilate below 35 MeV the baryon chemical potential quickly shoots up from ~1 eV to the nucleon mass (see ref.~\cite{Fromerth02} for more details). Effective models of QCD \cite{Scavenius01,Roessner07} as well as lattice calculations \cite{Fodor04} at finite baryon chemical potential give hints for the existence of a critical endpoint at $\mu_C = \mathcal{O}(1)T_C$. 

\begin{figure}[ht]
	\centering
		\includegraphics[width=0.48\textwidth]{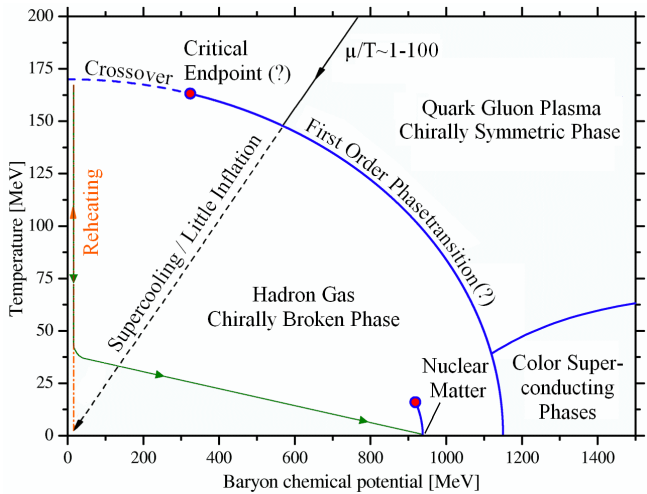}
		\caption{Sketch of a possible QCD-phase diagram with the evolution path of the universe in the little inflation scenario.}
	\label{fig:qcdpd}
\end{figure}

At this point one might ask if there is a simple scenario with the cosmological QCD phase transition being first order without violating the constraint of a small baryon asymmetry in the later evolution of the universe. In a recent publication we have introduced the little inflation scenario \cite{Boeckel10}, that allows for such a first order QCD phase transition in the early universe without being in contradiction to present cosmological observations. In figure \ref{fig:qcdpd} we sketch the evolution path of the universe in the little inflation scenario. Here the universe starts out at a large baryon chemical potential and therefore crosses the first order phase transition line but stays in the deconfined chirally symmetric phase. The universe is trapped in the wrong QCD vacuum state and undergoes a short period of inflation until the delayed phase transition takes place. The released latent heat then causes a large entropy release that dilutes the baryon asymmetry to the presently observed value. Afterwards the universe evolves along the standard path just as in figure \ref{fig:qcdpdstandard}.

The concept of a little inflation (or tepid inflation) at the QCD phase transition has been
also introduced earlier by K\"ampfer et al. \cite{Kaempfer86,Boiko90,Jenkovszky90,Kaempfer00} and for a inflationary period of similar duration as discussed here later by Borghini et al.~\cite{Borghini00}. In both cases an initially higher net baryon density is diluted to the presently observed small value of the baryon-to-photon ratio in the course of the QCD phase transition. The general idea of a short inflation to reduce a too high baryon asymmetry was mentioned even earlier by Linde in a publication on Affleck-Dine baryongensis \cite{Linde85} but not explicitly in the context of the QCD phase transition.

In the present manuscript we will reexamine the findings presented there in more detail and employ a chiral effective model of QCD to study some of the implications more thoroughly. The structure of the paper is as follows. In sec.II we address the first prerequisite of the scenario, namely a baryon asymmetry of order one before the QCD phase transition. In sec.~III we discuss the second prerequisite, i.e.~the topic of nucleation and especially the issues of supercooling and high surface tension. In sec.~IV we introduce the dilaton quark meson model and apply its results in sec.~V where we discuss the implications for linear structure formation. Sec.~VI deals with the changes to dark matter physics. Generation and modification of magnetic fields and gravitational waves are discussed in sec.~VII and VIII, respectively. In the appendix we summarize some analytic limits for the structure formation calculations that are mostly not found in the literature to our knowledge and may help with the understanding of the numerical results in sec.~V.

\section{baryon asymmetry}
\label{baryon asymmetry}
As we have seen one of the main requirements of such a short inflationary period at the 
QCD phase transition is a non-vanishing baryochemical potential $\mu_B / T\sim \mathcal{O}(1)$. Big bang nucleosynthesis calculations predict the observed primordial abundances of elements correctly only if the baryon asymmetry was tiny at a temperature of 1 MeV and below. The cosmic microwave background radiation as well as large scale structure observations predict very similar values and combining all these observations one finds a baryon asymmetry of $5.9\cdot10^{-10} < \eta_B < 6.4\cdot10^{-10}$ at $98\%$
confidence \cite{Steigman08}. 

Now we need to estimate how long such a little inflation has to be in order to start out with a sufficiently large ratio of $\mu_B/T$. The net number of baryons in a comoving volume is conserved and can be estimated by $N_B \approx a_i^3 \mu_{Bi} T_i^2 \simeq a_f^3 \mu_{Bf} T_f^2$ where the index $i$ refers to the initial values when
the vacuum energy starts to dominate the energy budget of the universe and $f$ to the final values after reheating. 
Therefore the initial ratio of the chemical potential to the temperature can be higher by
\begin{equation}
\frac{\mu_{Bi}}{T_i} \simeq \theta^3 \frac{\mu_{Bf}}{T_f} \left(\frac{T_f}{T_i}\right)^3 \label{muoT}
\end{equation}
with $\theta=a_f/a_i$. If the phase transition at the end of inflation transpires on a timescale much shorter than the Hubble time then the universe reheats back to the initial temperature at the start of inflation in good approximation $T_i\simeq T_f$. Then we can conclude from equation (\ref{muoT}) that for $\theta \sim 10^3 \approx e^7$  the baryon asymmetry before inflation $\eta_{Bi}$ and $\mu_i/T_i$ will be of order unity. The latter condition would, as we have seen, suffice to allow the QCD phase transition to be first order. 

The next question we need to address is if such a high initial baryon asymmetry is possible within one of the established baryogenesis mechanisms. Baryogenesis has been a long standing problem in cosmology ever since the pioneering publication by Sakharov \cite{Sakharov67} and is still a very active field of research especially since it became clear that successful baryogenesis requires physics beyond the standard model, see \cite{Dolgov92,Dine03,Buchmueller05,Buchmueller05B,Buchmueller07} for some extensive review articles of the field.\\ In the well established Affleck-Dine mechanism of baryogenesis \cite{AffleckDine85,Linde85} a large baryon asymmetry is much more natural than in other classes of models like for example baryogenesis via leptogenesis.
In short the idea is that baryon- and lepton-number carrying scalar fields with very flat potentials can locally aquire very large expectation values. The Affleck-Dine mechanism can readily be incorporated into supersymmetric models \cite{Dine96}, where squark- and slepton-fields play the role of the baryonic scalar fields. Once supersymmetry is broken the flat directions are lifted and the scalar-condensates decay to standard model particles leaving a finite baryon and lepton asymmetry. In simple realisations the Affleck-Dine mechanism can easily produce a too high baryon asymmetry for the standard cosmological scenario, thus either models with multiple fields or more sophisticated coupling terms have to be introduced to limit the initial baryon number production or a subsequent reduction is necessary. The latter could be achieved, as mentioned earlier, by a large entropy release that dilutes the baryon to photon ratio to the right value observed today for example by an inflationary period (see e.g.~ref.~\cite{Linde85}). That being said Affleck-Dine baryogenesis can provide $\eta_B \sim \mathcal{O}(1)$, where this is probably an upper limit \cite{Linde85}. Still, this bound has to our knowledge not been explored any further after the estimates in the initial publications by Affleck, Dine and Linde for the obvious reason that an even higher baryon asymmetry was not desireable.

Now we want to make the above rough guess for the highest possible $\mu_B$ before such a little inflationary period a bit more quantitative taking $\eta_B = 1$ as an upper limit. Due to asymtotic freedom QCD should be well described by a free gas of quarks and gluons at sufficiently high energy densities so we can use this to gain some more quantitative estimates. To keep things simple we take
all particles to be massless. The energy density, pressure, entropy density and number density of a relativistic gas read 
\begin{eqnarray}
\rho & = & g\left( \frac{\pi^2}{30} T^4 +  \frac{1}{7} \mu^2T^2 + \frac{1}{14 \pi^2} \mu^4 \right)\\
p & = &  \frac{\rho}{3} = g\left( \frac{\pi^2}{90} T^4 +  \frac{1}{21} \mu^2T^2 + \frac{1}{42 \pi^2} \mu^4 \right)\\
n & = & \frac{\partial p}{\partial \mu} = g\left(\frac{2}{21}T^2\mu+\frac{2}{21\pi^2}\mu^3\right) \label{numberdensity}\\
s & = & \frac{\rho + p -\mu n}{T} = g\left(\frac{2\pi^2}{45}T^3+\frac{2}{21}\mu^2 T\right)\label{entropydensity}
\end{eqnarray}
Here g is the effective number of bosonic helicity states, i.e. fermionic helicity states are weighted with a factor of $\frac{7}{8}$. For $\bar{n}_B$ we can directly use equation (\ref{numberdensity}) with $g = g_q / 3$ for the degrees of freedom. The entropy density has contributions from particles with sizable chemical potential and from those without, therefore we label the quark degrees of freedom an index $q$ and those that have a non-negligible chemical potential with an index $\mu$. This is necessary because both are not necessarily the same since leptons should most likely carry an asymmetry similar to the baryonic one.
\begin{eqnarray}
s & = & g\frac{2\pi^2}{45}T^3+g_\mu\frac{2}{21}\mu^2 T
\end{eqnarray}
If we now combine both we arrive at an estimate for the baryon asymmetry
 \begin{eqnarray}
\eta_B & = &\frac{\frac{2 g_q}{63}\left(T^2\mu+\frac{\mu^3}{\pi^2}\right) }{g\frac{2\pi^2}{45}T^3+g_\mu\frac{2}{21}\mu^2 T} = \frac{g_q 5\left(\frac{\mu}{T}+\frac{1}{\pi^2}\frac{\mu^3}{T^3}\right) } {g 7 \pi^2 + g_\mu 15\frac{\mu^2}{T^2}}\label{etaB}
\end{eqnarray}
Interestingly this means that in the limit of $\mu \gg T$ as well as in the limit $\mu \ll T$ the baryon asymmetry is just proportional to $\mu/T$. The limits can be directly read of to be
\begin{eqnarray}
\eta_B\approx \begin{cases}\frac{g_q 5}{g 7\pi^2} \frac{\mu}{T}\hspace{5mm} \mu \ll T \\ \frac{g_q}{g_\mu3\pi^2} \frac{\mu}{T} \hspace{5mm}  \mu \gg T \end{cases}\label{scalerule}
\end{eqnarray}
Here we assume for simplicity that all particle species with a non-zero chemical potential have the same chemical potential i.e. $\mu=\mu_q = \mu_\nu = \mu_e$ et cetera. In the end we assume that the equilibrium condition for the quark and baryon chemical potential holds $\mu_B = 3 \mu_q$. Note that one cannot treat baryons as fundamental degrees of freedom satisfying equation (\ref{numberdensity}) with a charge of $1/3$ within this simple estimate or there would be a contradiction to the chemical equilibrium condition.

\begin{figure}[ht]
	\centering
		\includegraphics[width=0.48\textwidth]{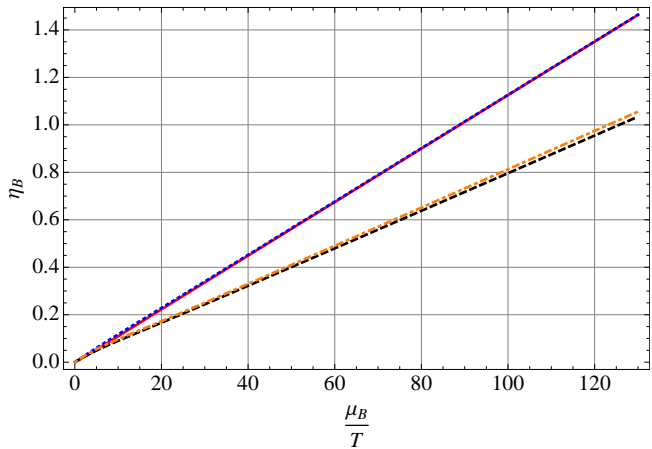}
		\caption{Here we plot the resulting baryon asymmetry from equation (\ref{etaB}) as a function of $\mu_B/T$ for different cases. All curves include photons, three neutrino families, electrons and positrons, up and down quarks and eight gluons. Solid red ($A$) and dotted blue ($B$) curves assume a negligible lepton asymmetry while the latter also includes strange quarks and muons. The dashed black ($C$) and the dashed-dotted orange ($D$) lines include a lepton asymmetry and again the latter adds s-quarks and muons.}
	\label{fig:etaBlin}
\end{figure}

In figure (\ref{fig:etaBlin}) the results from (\ref{etaB}) are shown for two particle compositions each for negligible lepton asymmetry and for equal baryon and lepton asymmetry. One can see that the influence from the particle composition is only a small effect, since the additional degrees of freedom contribute in a similar magnitude to numerator and denominator of $\eta_B$. On the other hand $\eta_B$ is significantly suppressed at the same chemical potential when adding an equal lepton asymmetry. This can be easily understood since the lepton asymmetry only increases the entropy but not the baryon number. The limiting values for $\mu_B/T$ assuming $\eta_B=1$ for the four cases are shown in table \ref{tab:etaB}

\begin{table}[h]
\centering
\begin{tabular}{ | c || c | c | c | c |}
\hline
  ~ & all particles & quarks & asym. particles  & $\mu_B/T|_{max}$  \\
  &$g$& $g_q$& $g_\mu$& for $\eta_B=1$\\\hline\hline
  A&47.75&21&21&88.89\\\hline
  B&61.75&31.5&31.5&88.74\\\hline
  C&47.75&21&29.75&125.7\\\hline
  D&61.75&31.5&43.75&123.1\\\hline
\end{tabular}
\caption{\label{tab:etaB}Degrees of freedom in the 4 considered cases A-D correspond to the curves in figure \ref{fig:etaBlin}}
\end{table}

Next we can translate the limits on the initial chemical potential to temperature ratio to a constraint on the extent of the dilution. The baryon number in a comoving volume is conserved, i.e.~$n_{Bi} = \theta^3 n_{Bf} $, therefore the extend of the dilution can be directly inferred from the ratio of baryon asymmetries before and after inflation
\begin{equation}
\theta = \left(\frac{\eta_{Bi}s_i}{\eta_{Bf}s_f}\right)^{1/3} = \left(\frac{\eta_{Bi}}{\eta_{Bf}}\right)^{1/3}\label{theta}
\end{equation}
Note that this definition does not necessarily coincide with the length of the period of exponential expansion defined by negative total pressure as we shall see later. To evaluate this expression we only need to calculate the baryon asymmetry $\eta_{Bi}$ because the two specific entropy densities $s_i$ and $s_f$ are by definition equal.

\begin{figure}[h]
	\centering
		\includegraphics[height=0.48\textwidth,angle=270]{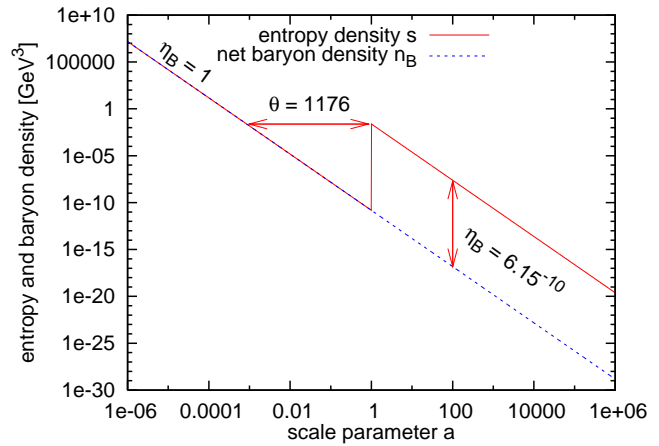}
		\caption{The reduction of $\eta_B$ to the presently observed value from an initial value of one. The scale parameter $a$ is normalized to the scale parameter at which reheating occurs.}
	\label{fig:entropy}
\end{figure}

If we now make use of the experimental value for $\eta_{Bf}$ we find the upper limit on the inflation length is given by
\begin{equation}
\theta_{max} = 1176~ \eta_{Bi}^{1/3}
\end{equation}
independent of particle composition. In figure \ref{fig:entropy} we show the corresponding maximum dilution of baryon number by a delayed QCD phase transition in the little inflation scenario. This figure is part of the results of the dilaton quark meson model and the structure formation calculation in sections IV and V, respectively, but it is quite model independent apart from the value of the chosen value of the vacuum energy.

The period of exponential expansion could also be estimated by comparision with the condition $p_V+p_R=0$, i.e. the point at which the pressure turns negative. Even such a simple estimate turns rather lengthy and would also not be very accurate because for interesting inflation lengths the dark matter energy density becomes of similar order than the vacuum and radiation energy densities. We will show the numerical results for the period of exponential expansion in contrast to the length of dilution as shown in figure \ref{fig:etaBlin} in section V.

\section{nucleation}
The next critical requirement of the little inflation scenario is a large supercooling or in other words if a sufficiently delayed phase transition is possible. This issue is directly connected to the stability and height of the barrier between the chirally broken phase and the chirally restored phase in the effective potential for sufficiently low temperatures. In chiral models of QCD including gluonic degrees of freedom in the form of a dilaton field the barrier only vanishes in the $T\rightarrow 0$ limit \cite{Campbell90} thus strong supercooling is in principle possible and we will come back to this model later on.

First let us consider the nucleation rate $\Gamma$ of the low temperature phase inside the high temperature phase
\begin{equation}
\Gamma = \Gamma_0 e^{-\Delta F_*/T}\label{nucleation rate}
\end{equation}
where the functional form is that of a thermally activated process as found by Langer in the 60s and 70s, e.g.~\cite{Langer69}. $\Gamma_0$ is in general a temperature dependent dynamical prefactor and $\Delta F_*$ is the free energy needed to produce a critical sized bubble of the new phase inside the old phase. What is meant by a critical sized bubble in this context? If the temperature is smaller than the critical temperature of the phase transition $T < T_c$  the system becomes metastable and statistical fluctuations produce bubbles of the low temperature phase with a radius $R$  and a free energy of
\begin{equation}
\Delta F = \frac{4\pi}{3}\left(p_H(T) - p_L(T)\right)R^3 + 4\pi R^2 \sigma_S
\end{equation}
Here $p_H(T)$ and $p_L(T)$ is the pressure in the high and the low temperature phase, respectively, and $\sigma_S$ is the surface tension. The first term describes the energy gained by transforming a spherical volume of radius $R$ to the new phase while the second term gives the energy it costs to create the surface interface around the bubble. Since $p_L(T)>p_H(T)$ both terms have opposite sign and there is a critical radius $R_*$ at which $\Delta F$ has a minimum
\begin{equation}
R_* = \frac{2\sigma_S}{p_L(T) - p_H(T)}
\end{equation}
only bubbles larger than $R_*$ can grow, for smaller ones it is energetically more favourable to shrink and disappear. One might just estimate $\Gamma_0$ by $T^4$ for dimensional reasons but Csernai and Kapusta \cite{Csernai92B} found $\Gamma_0$ in an effective field theory to be
\begin{equation}
\Gamma_0 = \frac{16}{3\pi} \left(\frac{\sigma_S}{3T}\right)^{3/2} \frac{\sigma_S \eta_H R_*}{\xi_H^4 (\Delta w)^2}
\end{equation}
which can easily be a few orders of magnitude smaller than the naive estimate. Here $\eta_H$ and $\xi_H$ are the shear viscosity and the correlation length in the high $T$ phase, respectively, and $\Delta w$ is the difference in enthalpy density $w = \rho + p $ between the two phases.

The important ratio for the cosmological QCD phase transition is $\Gamma/H$, i.e. the rate of nucleation to the Hubble parameter. Once this ratio exceeds unity bubbles are produced abundantly and coalesce until the transition is complete. If $\Gamma/H$ does not exceed one then bubbles of the low temperature phase will form and grow but the distance between bubbles increases so fast that the volume fraction of the new phase stays small. 

We will in the following compare to work done by Csernai and Kapusta \cite{Csernai92,Csernai92B} for the QCD phase transition within the bag model to find if the nucleation rate can be sufficiently small compared to the Hubble parameter such that the phase transition will initially fail. In ref.~\cite{Csernai92} the authors found that the  transition is completed very quickly with only marginal supercooling of about 1\% below the critical temperature. In fact this result depends strongly on the value of the surface tension $\sigma_S$ which they took to be $\sim 50 $MeV/fm$^2$. This number originates from an older work of Kajantie et.~al \cite{Kajantie90} who calculated the surface tension at critical temperature and zero density, for which the transition is found to be a crossover by all recent lattice calculations. As one can see from equation (\ref{nucleation rate}) $\Gamma$ depends exponentially on the value of the surface tension as well as on the free energy difference between both phases and especially the former quantity is in principle unknown at non-zero baryon density.

\begin{figure}[ht]
	\centering
		\includegraphics[width=0.48\textwidth]{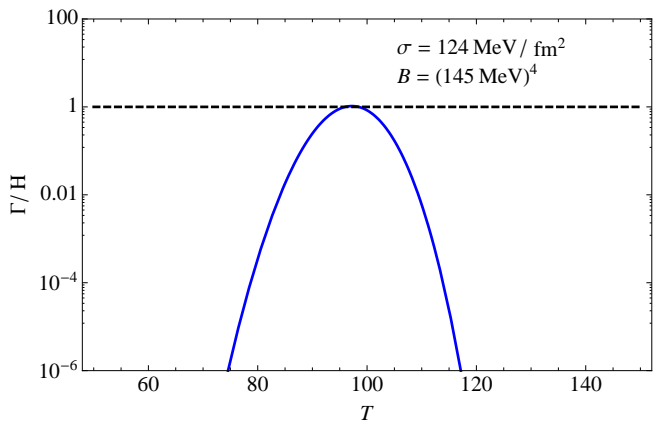}
		\caption{Nucleation rate over the Hubble parameter for the lowest value of the surface tension for which the phase transition would initially fail.}
	\label{fig:nucleation}
\end{figure}

\begin{figure}[ht]
	\centering
		\includegraphics[height=0.48\textwidth,angle=270]{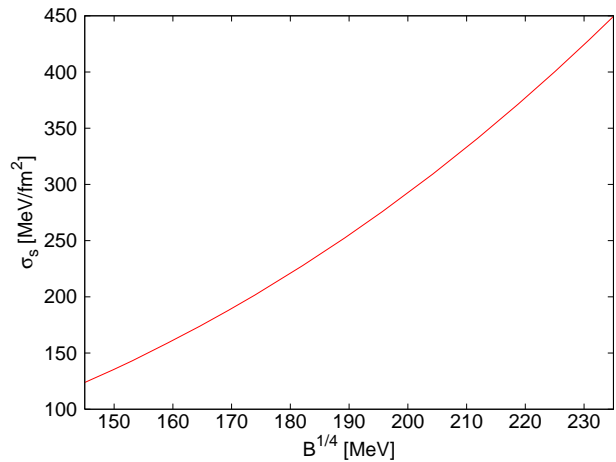}
		\caption{Minimum value of the surface tension as a function of the Bag constant at which $\Gamma/H$ does not exceed unity.}
	\label{fig:sur}
\end{figure}

Using the bag model with $T_c = 170 MeV$ and the same parameters as in \cite{Csernai92B} we looked for the lowest surface tension at which $\Gamma / H$ does not exceed unity at least until its maximum at around $\sim T_c / 2$. This might already be overstreching the applicability of (\ref{nucleation rate}) but it should still give a reasonable estimate of the surface tension needed for nucleation to fail. We find that the surface tension must indeed be very large and exceed $448~\mbox{MeV/fm}^2 \sim 3.7~T_c^3$ using their high value of the bag constant of $\mathcal{B} = ($235MeV$)^4$. If we however go to the lower end of values found in the literature, i.e.~ the original number $\mathcal{B} =  \left(145\mbox{MeV}\right)^4$  found by the MIT group to fit hadron masses \cite{Degrand75}, we find that a significantly lower $\sigma_S =$ 124 MeV/fm$^2$ suffices. The resulting  $\Gamma / H$ in that case is shown in figure \ref{fig:nucleation}. The surface tension for the QCD phase transition at non-zero baryon densities can only be estimated by effective models since lattice gauge theory calculations for this case are still in its infancies. In ref.~\cite{Voskresensky02} a reasonable range of $\sigma_S= 50 - 150$ MeV/fm$^2$ is discussed but even smaller or larger values are not excluded in principle. In Ref.~\cite{Palhares10} the surface tension was computed within the linear sigma model
to be as low as 5-15 MeV/fm$^2$. If one considers very high densities the surface tension for the transition from color superconducting phases to nuclear matter could reach values of 300 MeV/fm$^2$ \cite{Alford01}. In figure \ref{fig:sur} the minimal surface tension needed for nucleation to fail is shown for the commonly discussed range of the bag constant.

Our estimate only covers the initial failure to nucleate but it is clear that the phase transition has to occur after some limited supercooling (compared to ordinary inflation) of only about 7 e-foldings at most as we have seen. We stress that one should not take this estimate too far because both $\mathcal{B}$ and $\sigma_S$ have to be temperature dependent in general since both will in a field theoretical approach originate from the relative height and the shape of the barrier between the two minima in the effective potential. Finally $\Gamma / H$ must exceed unity for inflation to end and the phase transition to proceed, for which the surface tension has to drop sufficently fast such that fluctuations can easily overcome the barrier. Another possiblity would be the complete vanishing of the barrier and a spinodal decomposition as studied for example in \cite{Jenkovszky90} for a bag like model. Other authors have also discussed the strong sensitivity of nucleation rates for example in the context of neutron stars and core-collapse supernovae. There it was found that nucleation timescales can basically not be constrained and range from $\mu$s up to the age of the universe \cite{Mintz10}.

Also for heavy ion collisions strong supercooling is discussed for the "quench"-scenario, see e.g.\ \cite{Scavenius99}. There the chiral phase transition is delayed as the field is trapped in a metastable minimum and is only released to the true minimum in the T=0 limit.

The equation of state has to fulfill the usual condition $\rho + 3 p < 0$ to enter an inflationary phase. In the bag model this would be the case below a temperature $T_{inf}=\left(30 \mathcal{B}/(g\pi^2)\right)^{1/4}$. In the linear-$\sigma$-model or the NJL-model this occurs when the thermal contributions to the energy density become smaller than the vacuum contributions like the quark condensate $\left<m_q q \bar{q}\right> \approx f_\pi^2m_\pi^2$ and the gluon condensate $\beta_{QCD}/(2g) \left<G^a_{\mu\nu}G_a^{\mu\nu}\right> \approx 4\mathcal{B}$.

We can conclude that QCD at non-zero baryon densities is only poorly constrained and a delayed chiral phase transition is very well possible and has already been discussed for several other scenarios apart from the early universe.

\section{dilaton quark meson model}
\begin{figure}[htb]
	\centering
		\includegraphics[height=0.48\textwidth,angle=270]{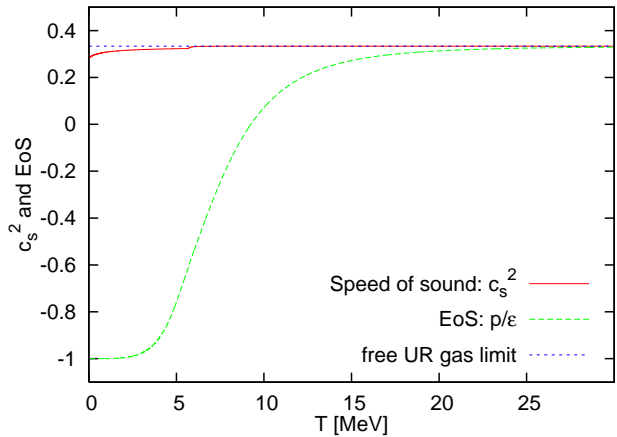}
		\caption{Square of the isentropic speed of sound and equation of state for the most extreme case $\mu_B/T = 125.7$.}
	\label{fig:eos}
\end{figure}

\begin{figure}[htb]
	\centering
		\includegraphics[height=0.48\textwidth,angle=270]{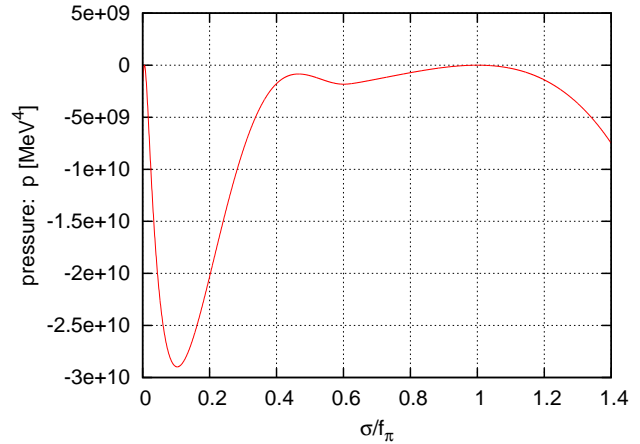}
		\caption{Pressure p as a function of the $\sigma-$field in units of $f_\pi$ at the phase transition temperature $T = 10.1$ MeV and $\mu_B/T=125.7$.}
	\label{fig:pressure}
\end{figure}

To describe the dynamics of the phase transition and especially the impact on density perturbations it is essential to have a reasonable thermodynamic description of the chirally restored quark phase. For this we use the quark meson model with a dilaton field, which incorporates chiral symmetry breaking as well as the trace anomaly of QCD. This model has been discussed by numerous authors \cite{Mishustin93,Heide94,Papazoglou97,Bonanno09} to describe nuclear matter. It has the interesting property that for a wide range of parameters the high temperature phase does only disappear in the $T\rightarrow0$ limit \cite{Campbell90}, which is necessary to get an equation of state of the "wrong vacuum" that can be used to model inflation. We will apply a simplified version of the lagrangian used in \cite{Papazoglou97} and stick closely to their notation, while we do not include the $\omega$-meson for simplicity and use quarks instead of nucleons as degrees of freedom. The Lagrangian includes the linear $\sigma$-model first introduced by \cite{Gell-Mann60} which incorporates the scalar isovector $\pi$-field with the $\sigma$-field as its chiral partner. Furthermore we include the isoscalar dilaton field $\chi$ that incorporates the scale anomaly and thus a non-trivial vacuum of QCD. The Lagrangian reads
\begin{eqnarray}
\mathcal{L} &=& \frac{1}{2}\left(\partial_\mu\pi\right)^2 + \frac{1}{2}\left(\partial_\mu\sigma\right)^2+\frac{1}{2}\left(\partial_\mu\chi\right)^2\nonumber\\&+& \bar\psi\left[i\slash{\!\!\!\partial}-g(\sigma+i\gamma_5\vec{\tau}\cdot\vec{\pi})\right]\psi -U(\pi,\sigma,\chi)
\end{eqnarray}
where the potential is given by 
\begin{eqnarray}
&&U(\pi,\sigma,\chi) = \frac{\lambda}{4}(\sigma^2+\pi^2)^2-\frac{k_0}{2}\left(\frac{\chi}{\chi_0}\right)^2(\sigma^2+\pi^2)\nonumber\\&-&f_\pi m_\pi^2\sigma \left(\frac{\chi}{\chi_0}\right)^2 + k_1\left(\frac{\chi}{\chi_0}\right)^4+ \frac{1}{4}\chi^4\ln\frac{\chi^4}{\chi_0^4}
\end{eqnarray}
Note that the choice of the term $-f_\pi m_\pi^2\sigma \left(\frac{\chi}{\chi_0}\right)^2$ that breaks chiral symmetry explicitly is not unambigous. The given choice $\chi^2/\chi_0^2$ corresponds to a fermion mass term, while $\chi/\chi_0$ would lead to a bosonic mass term. The differences caused by this choice are small since the symmetry breaking term logarithmic in $\chi$ is usually much larger, but for a similar model including vector mesons the quadratic choice is favored when comparing to nuclear matter \cite{Heide94}.\\
After integrating out the quark degrees of freedom \cite{Mocsy04} one arrives at an effective mesonic lagrangian 
\begin{eqnarray}
L(\pi,\sigma,\chi)  &=&  \frac{1}{2}\left(\partial_\mu\pi\right)^2 + \frac{1}{2}\left(\partial_\mu\sigma\right)^2+\frac{1}{2}\left(\partial_\mu\chi\right)^2\nonumber\\ &-& U(\pi,\sigma,\chi) - \Omega_{\bar q q} (T,\mu,m_q)
\end{eqnarray}
with the quark-antiquark potential reading
\begin{eqnarray}
\Omega_{\bar q q} (T,\mu,\sigma) = &-&\frac{\nu_qTV}{2\pi^2}\int^{\infty}_{0} dp p^2 \left[\ln\left(1+e^{-\beta (E_q-\mu)}\right)\right. \nonumber\\&+& \left.\ln\left(1+e^{-\beta (E_q+\mu)}\right) \right]
\end{eqnarray}
Here the single particle energy is as usual given by $E_q=\sqrt{p^2+m_q^2}$. The effective quark mass is on the mean field level determined by $\label{mqsimple}m^2_q = g^2 \sigma^2$.
The glueball and meson fields have the following mean values in the vacuum $\left<\chi\right>=\chi_0, \left<\sigma\right>=\sigma_0 = f_\pi$ and $\left<\pi\right>=0$. The full thermodynamic potential is then given by
\begin{equation}
\Omega(T,\mu,\pi,\sigma,\chi) = (U(\pi,\sigma,\chi) - U_{vac})V + \Omega_{\bar q q} (T,\mu,\sigma)
\end{equation}
Where $U_{vac}$ is subtracted to ensure the correct normalization $\Omega(0,0,0,f_\pi,\chi_0) = 0$, yielding
\begin{equation}
U_{vac} = \frac{\lambda}{4}f_\pi^4-\frac{k_0}{2}f_\pi^2-f_\pi^2 m_\pi^2+ k_1
\end{equation}
$\Omega/V$ exhibits two minima, one at $\sigma\sim0$ and $\chi\sim0.8\chi_0$ and a second one at $\sigma\sim f_\pi$ and $\chi\sim\chi_0$. The former corresponds to the chirally restored phase with a low effective mass while the second one is the chirally broken phase with a large effective mass. Full restoration of scale symmetry, i.e. the first minimum being located at $\sigma\sim0$ and $\chi\sim0$, is only realized at high temperatures and low densities at least for the parameters considered in our calculation. If the scalar coupling is sufficiently large both minima are present in the low temperature limit, although the chirally broken phase is energetically favoured. As the authors of Ref.~\cite{Bonanno09} have found the chirally restored phase will undergo a crossover to restored scale symmetry at much higher temperatures if the density is non-zero, i.e. the maximum moves towards $\chi\sim0$.
Note that we do not fix the effective quark mass in the vacuum via the Goldberger-Treiman relation because the model is set up to describe quarks in the high temperature chirally restored phase. We will later on fix the model parameters using the more relevant vacuum energy constraints and the masses of the sigma meson and the dilaton. The constants $k_0$ and $k_1$ are determined by the conditions
\begin{eqnarray}
 \left.\frac{\partial\Omega/V}{\partial\sigma}\right|_{vac} = \left.\frac{\partial\Omega/V}{\partial\chi}\right|_{vac} = 0
\end{eqnarray}
The equations of motion are found by minimizing the thermodynamic potential with respect to $\sigma$ and $\chi$.
\begin{eqnarray}
\frac{\partial\Omega/V}{\partial\sigma} = 0 &=& \lambda \sigma^3-k_0\left(\frac{\chi}{\chi_0}\right)^2\sigma-f_\pi m_\pi^2 \left(\frac{\chi}{\chi_0}\right)^2\nonumber\\&+&g \rho_S\\
\frac{\partial\Omega/V}{\partial\chi} = 0& = &-k_0\frac{\chi}{\chi_0^2}\sigma^2 - 2 f_\pi m_\pi^2\sigma \frac{\chi}{\chi_0^2}\nonumber\\&+& \chi^3\left(\frac{4 k_1}{\chi_0^4}+1+\ln\frac{\chi^4}{\chi_0^4}\right) 
\end{eqnarray}
These equations can reduced to a one dimensional problem by solving for $\chi$ explicitly
\begin{equation}
\chi = \chi_0 \left(\frac{\lambda\sigma^3 + g \rho_S}{k_0\sigma+m_\pi^2f_\pi}\right)^{1/2}
\end{equation}
here $\rho_S$, the scalar density,  is defined by 
\begin{equation}
\rho_S = \frac{g \nu_q}{2\pi^2} \int_0^\infty dp p^2\frac{m_q}{E_q}\left[\frac{1}{e^{\beta (E_q-\mu)}+1} +\frac{1}{e^{\beta (E_q+\mu)}+1} \right] 
\end{equation}
The pressure is as usual just given by
\begin{equation}
P(T,\mu) = -\frac{\Omega}{V}
\end{equation}
the net quark density is calculated via
\begin{eqnarray}
&&\bar n_{q} (T,\mu,m_q) = \frac{\partial \Omega/V}{\partial \mu}\\\nonumber &=& \frac{\nu_q}{2\pi^2}\int^{\infty}_{0} dp p^2 \left[\frac{1}{e^{\beta (E_q-\mu)}+1}-\frac{1}{e^{\beta (E_q+\mu)}+1} \right]
\end{eqnarray}
The energy density is then given by
\begin{eqnarray}
&&\epsilon(T,\mu) = \left(1-T\frac{\partial}{\partial T}-\mu\frac{\partial}{\partial \mu}\right)\frac{\Omega}{V}\\
&&=U(\pi,\sigma,\chi) - U_{vac}\nonumber\\&+&\frac{\nu_q}{2\pi^2}\int^{\infty}_{0} dp p^2 E_q \left[\frac{1}{e^{\beta (E_q-\mu)}+1} +\frac{1}{e^{\beta (E_q+\mu)}+1} \right] \nonumber
\end{eqnarray}
The entropy density can as usual be deduced from the Euler-equation
\begin{equation}
\epsilon = T s - p V +\mu \bar n_q ~~~ \rightarrow ~~~ s = \frac{\epsilon+p-\mu \bar n_q}{T} 
\end{equation}
Calculating the speed of sound is a bit more involved but straightforward. By definition the speed of sound is the isentropic derivative of the pressure with respect to the energy density
\begin{equation}
c_s^2=\left.\frac{\partial p}{\partial \epsilon}\right|_s
\end{equation}
Isentropic means nothing else but
\begin{equation}
ds = \left.\frac{\partial s}{\partial T}\right|_\mu dT +\left.\frac{\partial s}{\partial \mu}\right|_T d\mu = 0 \label{isentropic}
\end{equation}
Using the total differentials of pressure and energy density and equation (\ref{isentropic}) we arrive at the speed of sound in the form
\begin{equation}
\left.\frac{\partial p}{\partial \epsilon}\right|_s = \frac{ s \left.\frac{\partial s}{\partial \mu}\right|_T -  \bar n_q \left.\frac{\partial s}{\partial T}\right|_\mu }{\left.\frac{\partial \epsilon}{\partial T}\right|_\mu \left.\frac{\partial s}{\partial \mu}\right|_T - \left.\frac{\partial \epsilon}{\partial \mu}\right|_T \left.\frac{\partial s}{\partial T}\right|_\mu} 
\label{SOSeq1}
\end{equation}
The remaining parameters $\chi_0$ and $\lambda$ are fixed via the QCD vacuum energy and the mass of the sigma meson. Investigating the QCD trace anomaly Ref.~\cite{Collins77} found that at tree level the trace anomaly of QCD is given by
\begin{equation}
\Theta^\mu_\mu=\beta_{QCD}/(2g) \left<G^a_{\mu\nu}G_a^{\mu\nu}\right>
\end{equation}
where $\Theta^{\mu\nu}$ is the energy momentum tensor, $\beta_{QCD}$ is the beta-function of QCD, $g$ is the strong coupling constant and $G^a_{\mu\nu}$ is the gluon field strength tensor.
The trace anomaly of QCD then relates the vacuum energy to the parameter $\chi_0$ \cite{Heide94}
\begin{eqnarray}
\Theta^\mu_\mu &=& 4 U - \sum_{\phi=\pi,\sigma,\chi}\phi\frac{\partial U}{\partial \phi}\nonumber\\
&=& f_\pi m_\pi^2\sigma + \chi^4 \simeq  \chi^4 =  4 \epsilon_{vac} \left(\frac{\chi}{\chi_0}\right)^4\nonumber\\
\rightarrow \epsilon_{vac} & = & \frac{\chi_0^4}{4}  \label{Evaclimit}
\end{eqnarray}
We neglect the contribution from the first term representing the quark condensate for simplicity as done in \cite{Papazoglou97} because its contribution is much smaller than the one from the gluon condensate given by the second term for the parameters we will use later on. QCD sum rules suggest $|\epsilon_{vac}| \approx (240$ MeV$)^4$ (see ref.~\cite{Vainshtein78}) while bag model estimates range from $ (235$ MeV$)^4$ down to  $(145$ MeV$)^4$ in the original paper of the MIT group \cite{Degrand75}. This results in a possible range for the parameter $\chi_0$ of $205$ MeV $ < \chi_0 < 339$ MeV. We choose a $\sigma$-mass of 642 MeV, a dilaton mass of 1.5 GeV and a vacuum energy of $(236$ MeV$)^4$ to achieve a critical temperature of $170$ MeV for the phase transition at zero net density. The scalar coupling is chosen to be g = 7.5 which is the limiting value above which the chirally restored phase is present even in the $T\rightarrow0$ limit approximately.\\
In figure \ref{fig:eos} we show the resulting speed of sound and equation of state in the maximum case $\mu_{Bi}/T_i = 125.7$
As one could expect the speed of sound stays very close to the relativistic gas value of $c_s^2=1/3$ because the effective quark mass stays low in the chirally restored phase. The equation of state nicely interpolates from a relativistic gas ($w=1/3$) to that of vacuum energy ($w=-1$). The small kink in the speed of sound is caused by the merging of a third always metastable intermediate phase with the chirally restored phase which causes a sudden but small change in the effective mass. The existence of this third maximum in the pressure within this model has been discussed before for example by Mishustinet al.~\cite{Mishustin93}, it can also be seen in figure \ref{fig:pressure} at $\sigma\sim 0.5 f_\pi$. There we show the pressure as a function of the $\sigma$-field (or equivalently the effective mass) at the phase transition temperature $T = 10.1$ MeV and $\mu_B/T=125.7$. The first minimum at $\sigma\sim0$ is the chirally restored phase, the chirally broken phase is located at $\sigma\sim f_\pi$. The intermediate phase only appears close to the phase transition and never becomes the favoured one. Note that at these temperatures and densities one may expect color-superconducting quark matter in one of many possible phases \cite{Alford08} which exceeds the scope of the current investigation but may be an interesting starting point for an alternative field theoretical description of the scenario.

In the next section we will use the $\epsilon, p , c_s^2$ and $w$ of the chially restored phase for our structure formation calculations.


\section{structure formation}
\noindent Next we will investigate the effect of a little inflationary period on primordial density perturbations. In particular dark matter perturbations are affected in several ways and on much larger scales than usual for the cosmological QCD phase transition. First of all the Hubble radius is roughly given by $R_H\sim g^{-1/2}m_{Pl} T^{-2}_c\sim 10$ km which encloses a total energy corresponding to about $1M_\odot$. Since this epoch is long before matter radiation equality, i.e. $\rho_{DM}\sim(a_{QCD}/a_{EQ})\rho_R\sim10^{-8}\rho_R$, the mass of dark matter in the same volume is smaller by the same factor resulting in a dark matter mass scale of approximatly $10^{-8} M_\odot$. About ten years ago Schmidt, Schwarz and Widerin investigated the effect of the QCD phase transition on dark matter perturbations \cite{Schmid97,Schmid99}. They found that peaks and dips in the spectrum of dark matter perturbations may form for a first order phase transition but even for a crossover one could expect a boost for small scale perturbations. These effects were due to the reduction of the speed of sound $c_s$ and equation of state $w=p/\rho$ of the radiation fluid during the phase transition. As the above estimate implies they only found these effects at very small mass scales below the Hubble scale.
We shall examine the little inflation scenario with the same approach to density fluctuations, i.e. we work in the so called uniform-expansion gauge (UEG) that is free of spurious gauge modes and well behaved in the superhorizon limit \cite{Bardeen80,Schmid99} which is especially important for the little inflation scenario since modes can enter the horizon and exit the horizon several times.

For the background evolution we assume a flat Friedmann-Robertson-Walker metric that implies the well known Friedmann equations
\begin{eqnarray}
H^2&=&\frac{8\pi G}{3}\rho\label{Hubble}\\
\dot H &=& -4\pi G (\rho + p)\label{DHubble}
\end{eqnarray}
The overdot as usual denotes a derivative with respect to time. The common approach now is to decompose the perturbed metric into scalar, vector and tensor perturbations where only scalar perturbations lead to the growth of structure. The corresponding perturbed line element reads
\begin{eqnarray}
ds^2= a^2(\eta) \left[ \left( 1+2\alpha \right) d\eta^2 - 2 B_{| i} dx^i d\eta\right. \nonumber\\\left. - \left( \left[ 1+ 2 \varphi \right] \gamma_{ij} + 2E_{| ij} \right) dx^i dx^j \right]
\end{eqnarray}
Here $\alpha,B,\varphi,E$ are spacetime dependent scalar functions representing the four scalar degrees of freedom and $\gamma_{ij}$ is the spacial part of the background metric. Following \cite{Hwang93} we introduce two combinations of the metric variables that are independent of spacial gauge transformations just like $\alpha$ and $\varphi$. These are
\begin{eqnarray}
\chi &\equiv& -a\left(B - H a^2 E^\prime \right) \\
\kappa &\equiv& 3H\left(\alpha+a\varphi^\prime\right) + \frac{k^2}{a^2}\chi 
\end{eqnarray}
where primes denote derivatives with respect to the scale parameter $a$ and $k$ is a comoving wavenumber. One can show that $\chi$ and $\kappa$ describe the perturbations in the shear $\sigma_{ij}$ and the expansion $\Theta$, respectively. The latter is given by $\Theta = 3 H - \kappa$, therefore choosing $\kappa=0$ corresponds to having an unperturbed Hubble flow. This explains the naming uniform expansion gauge or Hubble constant gauge. Choosing $\chi=0$ leads to the more popular longitudinal or conformal Newtonian gauge as used in the well known review of Mukhanov, Feldman and Brandenberger \cite{Mukhanov92}. For ideal fluids the evolution equations in UEG read \cite{Hwang93,Schmid99}
\begin{eqnarray}
\dot \epsilon &=& - 3 H (\epsilon + \bar\pi)-\Delta\psi-3H(\rho+p)\alpha\label{pert1}\\
\dot\psi &=& - 3 H \psi-\bar\pi-(\rho+p)\alpha\label{pert2}
\end{eqnarray}
which can be deduced from energy-momentum conservation and the three divergence of the Euler equation. Here $\epsilon \equiv \delta\rho$ and $\bar\pi\equiv \delta p$ denote the perturbation of the energy density and pressure, respectively. Furthermore $\psi$ is the potential of the momentum density $\vec S$, i.e.~$\vec\nabla\psi = \vec S$. The latter is related to the fluid velocity $v$ via $\psi k/a = (\rho + p) v$. Equations (\ref{pert1}) and (\ref{pert2}) apply for each decoupled ideal fluid. All fluids are gravitationally linked via the perturbation of the lapse $\alpha$ and Einsteins $R^0_0$-equation
\begin{eqnarray}
(\Delta+3 \dot H)\alpha=4\pi G (\rho+3p)\label{poisson}
\end{eqnarray}
Here one already realizes that $H^2$ and $\dot H$ appear as two distinct scales in the set of perturbation equations. These will be similar for most cases, but during an inflationary period they are not as we will see later. Introducing dimensionless variables $\delta=\delta\rho/\rho$, $\hat\psi= k\psi/(a\rho)$ and the equation of state $w=p/\rho$ the UEG set of equations takes the form
\begin{eqnarray}
\delta_i^\prime & = & - \frac{3(c_{si}^2-w_i)}{a}\delta_i + \frac{k}{\mathcal{H}a} \hat\psi_i - 3 (1+w_i) \frac{\alpha}{a}\label{density contrast}\\
\hat\psi_i^\prime & = & - \frac{1-3w_i}{a}\hat\psi_i - c_{si}^2 \frac{k}{\mathcal{H}a} \delta_i - (1+w_i) \frac{k}{\mathcal{H}a}\alpha\label{velocity}\\
\alpha & = & - \frac{\frac{3}{2}\left(1+ 3 c_{s}^2\right)}{\left(\frac{k}{\mathcal{H}}\right)^2+ \frac{9}{2}\left(1+w\right)}~\delta\label{alpha}
\end{eqnarray}
Here the index $i$ refers to an individual fluid, each of which has a set of equations (\ref{density contrast}) and (\ref{velocity}). $\mathcal{H} = H a$ is the so called conformal Hubble parameter. Primes denote derivatives with respect to the scale parameter. All fluids are connected via the last equation for the perturbation of the lapse $\alpha$. The mean density contrast, equation of state and speed of sound are calculated by
\begin{eqnarray}
\delta = \frac{\sum_i \delta_i \rho_i}{\sum_i\rho_i}, ~ w = \frac{\sum_i p_i}{\sum_i\rho_i}, ~ c_s^2 = \frac{\sum_i c_{si}^2 \delta_i \rho_i}{\sum_i \delta_i\rho_i}
\end{eqnarray}
Basically all viable dark matter candidates are already chemically decoupled
from the radiation fluid at the QCD phase transition, thus their numbers are not repopulated by reheating after inflation. Therefore the dark matter number density is diluted by the same factor
$\theta^3$ as the net baryon number. As stated before the dark matter mass
enclosed inside the Hubble horizon is of the order of $10^{-8}
M_{\astrosun}$ at $T_{QCD}\sim 170$ MeV. Thus any influence on
perturbations inside dark matter would not have any consequences on
larger scales. An inflationary period at the QCD-phase transition can
change this in two ways. First of all the amount of dark matter enclosed
inside the horizon must be larger by a factor $\theta^3$ initially to
match the present day dark matter density despite the dilution. For a short inflationary
period, as discussed here, one encounters an additional effect on
perturbations that have physical wavenumbers $k_{ph} \lesssim H$ at
the beginning of inflation. This is caused by an additional scale apart from $H$, namely $\dot H^{1/2}$, via equation (\ref{poisson}) emerges. One may realize this by combining equations (\ref{Hubble}) and (\ref{DHubble}) to find that
\begin{equation}
\frac{\dot H}{H^2} = -\frac{2}{3}\frac{\rho+p}{\rho} =  -\frac{2}{3}\left(1+w\right)
\end{equation}
So as long as $w$ is not too close to -1 both scales coincide, but during an inflationary phase this is no longer true. Let us do an estimate for a general mix of radiation, dark matter and vacuum energy. In this case
\begin{eqnarray}
\dot H = -4 \pi G \left[ \frac{4}{3}\rho_{Ri}\left(\frac{a_i}{a}\right)^4 +
\rho_{Mi}\left(\frac{a_i}{a}\right)^3 \right] \propto \left(\frac{a_i}{a}\right)^q
\end{eqnarray}
where the subscripts refer to matter and radiation with $q=3$ to 4,
respectively. The index $i$ refers to the onset of inflation. Comparing
this to the first Friedmann equation one finds that 
\begin{equation}H^2 = \frac{8\pi
  G}{3}\left[\rho_V +
  \rho_{Ri}\left(\frac{a_i}{a}\right)^4 + 
  \rho_{Mi}\left(\frac{a_i}{a}\right)^3\right] \approx \frac{8\pi
  G}{3}\rho_V
\end{equation}
\noindent As a consequence the two scales differ by 
\begin{equation}
\left|\frac{\dot H}{H^2}\right|^{1/2} \simeq \left(\frac{a_i}{a}\right)^{q/2},
\end{equation}
This would not play any role for a long inflationary period, i.e.~with more than 50 e-foldings. In this case $\dot H^{-1/2}$ is beyond the size of the observable universe, approximately at the order of the infrared cutoff of the produced primordial spectrum. Summarizing, there should be two distinct scales in the spectrum dividing it into three regimes
\begin{eqnarray}
\left.\frac{k_{ph}}{H}\right|_i &>& 1~~~~~~~~~~~~~ \mbox{(sub-hubble before inflation)}\nonumber\\
1 > \left.\frac{k_{ph}}{H}\right|_i &>& \left(\frac{a_i}{a_f}\right)^{q/2}~~~ \mbox{(intermediate)}\nonumber\\
 \left.\frac{k_{ph}}{H}\right|_i &<& \left(\frac{a_i}{a_f}\right)^{q/2}  ~~~\mbox{(unaffected)}\nonumber
\end{eqnarray}
One could even expect another spectral region of modes that are always sub-Hubble till the end of inflation. These would be located below $10^{-8} M_{\astrosun}$ and are thus of little relevance for structure formation.\\ 
Translating our previous estimates to the highest affected mass scale involved we find 
\begin{equation}
M_{max}\sim 10^{-8} M_{\astrosun}~\theta^{3q/2} \sim (10^{5}-10^{9}) M_{\astrosun}\label{mmax}
\end{equation}
at most for $\theta^{inf}\sim640$. In this case the second mass scale between the first and the second spectral region could be expected at 
\begin{equation}
M_{med}\sim 10^{-8} M_{\astrosun}~\theta^{3} \sim (1-10) M_{\astrosun}\label{mmed}
\end{equation}
\noindent For the numerical treatment we assume a scale invariant primordial Harrison-Zeldovich spectrum to be present before the phase transition. Each wavenumber k is followed separately from a point where is was sufficiently super-hubble to apply the initial conditions for a radiation dominated universe given by the growing super-horizon modes \cite{Hwang93,Schmid99}
\begin{eqnarray}
\delta_R = \frac{A}{6}\left(\frac{k}{\mathcal{H}}\right)^2 &~~& \hat\psi_R = \frac{A}{54}\left(\frac{k}{\mathcal{H}}\right)^3 \\ \delta_{DM} = \frac{3}{4}\delta_R &~~& \hat\psi_{DM} = \frac{9}{8}\hat\psi_R
\end{eqnarray}
\noindent For the description of the radiation background we take the input from the dilaton-quark-meson model as described in the previous section. We add a massless ideal gas of photons, gluons, $e^\pm$ and three neutrino families as in case C described in section \ref{baryon asymmetry}. For the dark matter we assume a decoupled pressureless non-relativistic gas with a vanishing speed of sound.
 
\begin{figure}[ht]
	\centering
		\includegraphics[height=0.50\textwidth,angle=270]{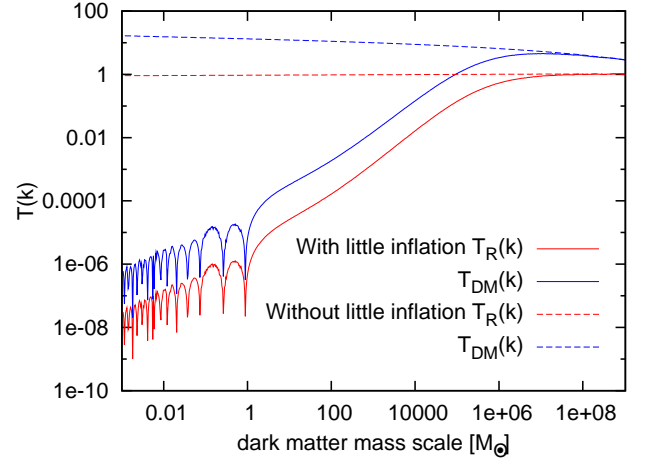}
		\caption{Spectrum of primordial fluctuations dependent on the dark matter mass scale in units of solar masses normalized to the amplitude at horizon entry. As input we used the equation of state of the chirally restored phase of a dilaton quark meson model with equal lepton and baryon chemical potentials (case C). For comparison the spectra without a little inflationary phase is shown with dashed lines.}
	\label{fig:spectrum}
\end{figure}

\noindent In figure \ref{fig:spectrum} we show the resulting spectrum of primordial fluctuations after a little inflation in comparison to the spectrum as expected without little inflation. The spectrum is given in terms of the transfer functions defined in the following way
\begin{eqnarray}
T_{R}(k) & = & \left(\frac{\delta^2_{R}(k)+\hat\psi_R^2(k)/c^{2}_{sR}}{\left(\delta_{R}^2+\hat\psi_R^2/c^{2}_{sR}\right)_{in}}\right)^{1/2}\\ 
T_{DM}(k) & = & \left(\frac{\delta^2_{DM}(k)}{\delta_{DM,in}^2}\right)^{1/2}
\end{eqnarray}
The "$in$" quantities are evaluated at (final) horizon entry in the limit of small wavenumbers \cite{Schmid99}, i.e.~in the unaffected part of the spectrum for the little inflation calculation. Note that a transfer function of unity does not correspond to non-linear perturbations. For a scale invariant spectrum it simply corresponds to a amplitude of $10^{-5}-10^{-4}$ since the perturbations in radiation are frozen until the decoupling of the cosmic microwave background radiation (ignoring low scale damping effects). The fluctuations are evolved for 12 orders of magnitude in $a$ to ensure that the whole spectrum is super-horizon at the start of the calculation and is completely sub-horizon at the end. The final temperature for the parameters used is $T_{end}\sim 150 $eV $\approx T_{reheat}/10^6$.\\
\noindent All scales below $M_{max}\sim 10^6 M_\odot$ show a suppression, those below $M_H\sim M_\odot$ show additional features depending on their phase during horizon exit. Above this scale the spectrum of density perturbations is given by the primordial spectrum of density perturbations, e.g.~a nearly scale invariant spectrum. In the appendix we also give the approximate analytic solutions for the different regimes. Summarizing they show that in the unaffected region the modes grow similarly to the radiation dominated super-horizon solutions while the intermediate modes are frozen. This consequently explains the relative suppression of the intermediate modes compared to the large scale limit.

\begin{figure}[ht]
	\centering
		\includegraphics[height=0.48\textwidth,angle=270]{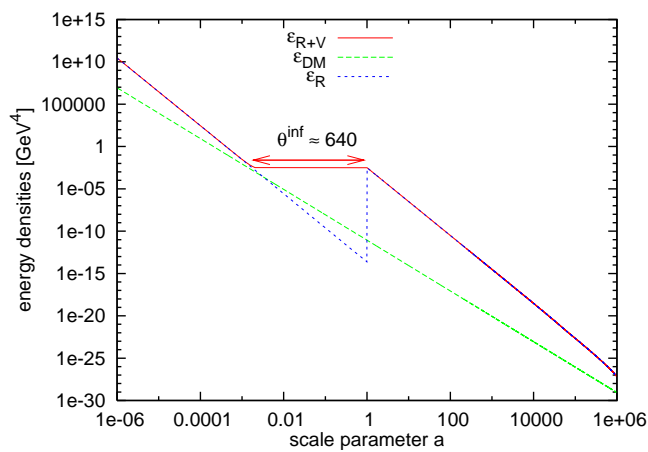}
		\caption{Evolution of the background energy densities of radiation plus vacuum contributions, dark matter and of radiation alone. As one can see the inflation length is only $\theta^{inf}\sim640$ while the dilution factor is $\theta = 1176$}
	\label{fig:densities}
\end{figure}

\noindent The numerical result for the maximum mass scale is quite close to the lower bound in aboves estimate (\ref{mmax}) because dark matter has to be more abundant than radiation during almost  the complete duration of the little inflation, which can be seen in figure \ref{fig:densities}. Still this mass scale is of cosmological interest as it is comparable to that of globular clusters (GC) which were the first objects to form during primordial galaxy formation (for a comprehensible overview of the topic the reader may have a look at the review by Harris \cite{Harris91}). Globular clusters are very compact star clusters of several hundred thousand to several million stars, with a radius of only $\sim 10$ pc. They are very metal poor objects and age estimates from stellar evolution models strongly suggest that they should already have been created during the formation of their host galaxy. Their mass function has a well defined peak at $M_{gc}\sim2\cdot 10^5 M_\odot$ in contrast to younger star clusters whose mass function shows a steep power law distribution between $10^4 M_\odot \lesssim M_{yc}\lesssim 10^7 M_\odot$ with an index of $\approx -2$ \cite{Prieto08}. There have been attempts to explain the preferred mass scale for GC by a higher Jeans mass at low metallicity that preferred more massive clusters at early times. Other explanations include disruptive processes (for low mass clusters) and mass loss due to stellar evolution (for high mass cluster) that might produce a preferred mass scale as seen in n-body simulations \cite{Fall01} starting from a steep mass function like the one of present young clusters. The latter point was dismissed by Vesperini et al.~\cite{Vesperini03} whose simulations have shown that only very fine tuned initial conditions will result in GC properties that fit the observations if one assumes GC formed just like clusters do today. On the other hand they showed that an initial mass function that was almost flat below the present peak mass scale succeeds in reproducing the observational data. Interestingly a power law cutoff in the dark matter fluctuations below a mass scale $10^{5-6} M_\odot$ as found in the little inflation scenario could thus help to explain this standing problem in galactic astrophysics.\\
The suppression of power spectrum below these scales could also be interesting to study because of its impact on the cuspy core density distribution of dark matter in small galaxies and the large number of halo structures seen in standard structure formation. 
\section{dark matter}
\noindent Apart from the impact on small scale structure formation for dark matter the little inflation scenario has further direct consequences on properties of dark matter candidates. For cold dark matter the dilution of the energy and number densities leads to the possibility of a matter dominated phase before the
inflationary phase since the dark matter energy density after reheating is basically fixed by the present day value. This can also be seen in figure \ref{fig:densities} that displays the evolution of the different contributions to the energy density, wherein dark matter just starts to dominate right before the onset of inflation. To account for the different ratio of radiation and baryon densities before little inflation the dark matter density has to be larger by the same factor $\theta^3$ as the baryon density. For $\theta \gtrsim 10^3$ the dark matter contribution actually becomes larger than the vacuum contribution which would in any case limit the maximum length of the exponential expansion to
\begin{eqnarray}
\theta^{inf}_{max} & = & \left(\frac{\mathcal{B}}{\rho_{DM}(a_f)}\right)^{1/3}
\nonumber\\ ~& \approx & 900 \left(\frac{\mathcal{B}^{1/4}}{235\mbox{MeV}}\right)^{4/3}
\left(\frac{0.236}{\Omega_{DM0}}\right)^{1/3}
\end{eqnarray}
where the bag constant $\mathcal{B}$ represents the vacuum contributions of QCD. Interestingly this limit and the previously discussed limit from the Affleck-Dine baryogenesis coincide by chance, while the latter actually limits the entropy release the former only limits the length of exponential expansion. As a side remark, to produce a complete spectrum of primordial fluctuations one would require $\theta \gtrsim 10^{10}$, far beyond both limits so little inflation cannot replace standard inflation.\\
Still figure \ref{fig:densities} shows that the period of exponential expansion $\theta^{inf}\approx 640$ is even shorter than this estimate because the energy density of radiation increases so strongly with the baryon asymmetry at a fixed vacuum energy. This difference in $\theta$ and $\theta^{inf}$ is caused by the different dependencies of the entropy and the energy density on $\mu$ or rather $n_B$.\\
 
\noindent What does a larger dark matter density before the QCD scale mean for the properties of cold dark matter? For non-relativistic decoupling of dark matter the weak interaction cross section will no longer give the right amount of dark
matter today. This is due to the fact that the dark matter annihilation cross section has to be
much smaller, i.e.
\begin{eqnarray}
\sigma^{annih}_{dm}\sim\frac{\sigma^{weak}}{\theta^3}~~~\mbox{because}~~~ \Omega_{DM}\propto\frac{1}{\sigma^{annih}_{dm}}\nonumber
\end{eqnarray}
where we ignore logarithmic dependencies on the dark matter mass. This allows more dark matter particles to survive annihilation before freeze-out and thus increases the CDM number density before little inflation. This gives the interesting prospect that the little inflation can be probed by ongoing and future collider experiments like the LHC since the discovery of a standard weakly interacting massive particle as the neutralino would exclude the scenario. 

\noindent Another case would be thermally decoupled ultra-relativistic particles where the dilution of dark matter number  densities can be incorporated in the ordinary temperature relation to the radiation background. Here little inflation leads to an effective shift in the temperature relation 
\begin{equation}T = T_{DM} \theta
\left(\frac{g^s_{eff}(T_{Dec})}{g^s_{eff}(T)}\right)^{1/3}
\end{equation}
\noindent This in turn modifies the relation of warm dark matter relic mass and decoupling degrees of freedom to match the present day density found for example in \cite{Boeckel07}
\begin{equation}
m^{max}_{DM} \approx 51 \mbox{eV} \theta^3
\left(\frac{4}{g_{DM}}\right)\left(\frac{g^s_{eff}(T_{Dec})}{106.75}\right)
\left(\frac{\Omega^0_{DM}h^2}{0.116}\right)
\end{equation}
\noindent This shifts the suitable mass of a thermal relic particle to a much higher value without the need for a large number of additional effective degrees of freedom at decoupling beyond
those of the standard model.\\
\noindent There can also be effects for baryonic dark matter as discussed by Jedamzik \cite{Jedamzik97}. During a first order phase transition the speed of sound vanishes and thus sufficiently nonlinear density fluctuations can collapse during that time. For an exponentially small fraction of Hubble volumes that are overdense enough primordial black holes (PBH) may form. The mass spectrum of these PBH will be strongly
peaked around $1 M_\odot$ which corresponds to the total (not just the
dark matter) energy density inside the Hubble volume at the phase
transition. The produced abundance of PBH depends on the spectral index and
amplitude of the density fluctuation spectrum, which we have seen is different and in general more complicated in the little inflation scenario. Nevertheless it seems quite clear that the suppression of small scale density fluctuations will also strongly reduce the production of such primordial back holes during the phase transition at the end of a little inflation. 

\noindent During the nucleation process lumps of quark matter or small quark stars could be produced but only with $M \sim 10^{-9} M_\odot$ as we argue that nucleation starts after the little inflationary epoch.\\

\section{magnetic fields}
A standing problem in astrophysics is the origin of large scale magnetic fields
that have strengths of up to $B^{obs}_\lambda=0.1 \mu$G on extragalactic and up to 10 $\mu$G on galactic
scales. To understand the existence of such magnetic fields with correlation lengths of typically 0.1 Mpc it is 
necessary to have an initial seed field generated before or during galaxy formation. The required strength of such seed fields varies strongly with the assumed amplification mechanism and may vary over many orders of
magnitude $10^{-30} G \lesssim B^{seed}_\lambda \lesssim10^{-10}$G, see \cite{Widrow02} and references therein for an overview of the topic.
The seed fields may be generated during ordinary inflation or at a first order phase transition. The latter has been
discussed for the QCD phase transition by numerous authors \cite{Witten84,Cheng94,Sigl97} at a time when the phase transition at small baryon-asymmetry was still believed to be first order. The established mechanism for magnetic field production was the collision of hadronic bubbles during the phase transition \cite{Cheng94}. Different masses of quarks and nucleons would lead to a diffusion of baryon number via the bubble walls and consequentially a baryon contrast close to the phase boundary would develope \cite{Witten84,Cheng94}. This baryon contrast can be estimated by the ratio of the net baryon numbers in the two phases to be
\begin{equation}
	\label{5}R=\frac{\bar{n}^B_q}{\bar{n}^B_H}
\end{equation}
Because muons and strange quarks are already slightly suppressed at the critical temperature $T_c$ the baryon contrast would also cause a charge dipole layer at the phase boundary to develope. The resulting net positive charge density is 
\begin{equation}
	\rho_C^+ =e\left(2/3 n_u - 1/3 n_d - 1/3 n_s\right) = \beta e n_B
\end{equation}
where the indices $u, d$ and $s$ refer to the different quark flavours and the factor of proportionlity $\beta$ depends on the temperature, chemical potential and the masses of the particles. For a small $\eta_B$ and reasonable strange quark and muon masses $\beta \sim 10^{-2}-10^{-3}$. After a strong supercooling muons and strange quarks will be suppressed resulting in $\beta\approx0.2$ for the little inflation case. Cheng et al.~estimated the magnetic field generated by the collision the hadron gas bubbles to be
\begin{equation}
	B_{QCD}\approx\frac{8\pi \rho_C \: r_d\: v}{3} = \frac{8\pi e R\: \beta\: \bar{n}_B\: r^2_{diff}\:H_{QCD}}{3}
\end{equation}
due to turbulent charged flow. Here the flow parallel to the bubble walls was assumed to have velocities $v \sim r_n H_{QCD}$ giving the main contribution to the field generation. The thickness of the the baryon excess layer $r_d$ was estimated according to results of \cite{Kurki-Suonio88} to be $r_d\approx r_{diff}^2/r_n$ with $r_{diff}$ being the baryon diffusion length and $r_n$ the mean separation of nucleation sites. $R\beta$ should be at least 0.3 with the above estimates up to values of $\sim 10-100$ if baryon number can be effectively piled up by the expanding bubble walls. Thus we arrive at magnetic fields of strength $B_{QCD}=10^8-10^{10}$G for low baryon asymmetry, i.e.~for the standart scenario assuming a first order phase transition. If the baryon contrast exceeds $R\sim10$ then any initial field may be readily amplified by magneto-hydrodynamic (MHD) turbulence  to the equipartition value (see \cite{Sigl97} and refs.~therein) 
\begin{equation}
	B_{eq}=\sqrt{8 \pi T^4 v_f^2}
\end{equation}
where $v_f$ is the fluid velocity. Now we shall modify these estimates for the little inflation scenario. First of all 
the initial value of the baryon number contrast $R$ between the two phases can be much higher because quarks are much more favourable carriers of baryon number than nucleons at such low temperatures of $T\sim 170\mbox{MeV}/\theta\sim 0.2$MeV at the end of inflation. The diffusion length will also be larger because both baryon and antibaryon densities $n_B, n_{\bar{B}}$ will be additionally diluted by a factor $\theta^3$ resulting in
\begin{equation}
	r_{diff} \propto 1/\sqrt{n_B+n_{\bar{B}}}\sim 4 \mu\mbox{m}~ \theta^{3/2} \sim 10~\mbox{cm}
\end{equation}
for a random walk approximation. Thus developement of MHD turbulence should be expected resulting in equipartition of the magnetic field with a strength of $B_{eq}\approx 10^{12}$G. The fluid velocities were taken to be $v_f\sim 1$ because the released latent heat is much larger than the thermal energy. 

Next one may ask if such a strong magnetic field does not violate bounds for the total the energy density foremost from big bang nucleosynthesis, which is the next important milestone in the evolution of the universe after the QCD phase transition. Caprini and Durrer found that magnetic fields produced by a causal production mechanism (in contrast to magnetic fields produced during primordial inflation) can be strongly limited via their integrated energy density and the shape of the spectrum \cite{Caprini02,Durrer03}. They argued that the spectrum of the generated magnetic field must fall off with a steep power law for uncorrelated superhorizon scales, i.e.\ $B^2_\lambda\propto \lambda^{-n}$ with $n \geq 2$. As stated earlier the typical comoving length scale of galactic magnetic fields is 0.1 Mpc which is comparable to the shortest magnetic field mode that survives plasma damping processies up to recombination \cite{Subramanian98,Jedamzik98}. This scale is clearly much larger than the comoving horizon size $H^{-1}\sim 10$ pc at the QCD phase transition. Therefore even a relatively small field strength at the 0.1 Mpc scale requires a magnetic field at the 10 pc scale that is larger by many orders of magnitude easily resulting in a very large integrated magnetic field energy density. We use the bound on an additional radiation energy density at big bang nucleosynthesis found by ref.~\cite{Cyburt05} allowing at most 1.6 additional effective neutrino families at the 98\% confidence level. The integrated magnetic energy density is thus bounded from above by $B_{QCD} = 5\cdot 10^{13}$G  which limits the strength of the comoving seed field to $B^{seed}_{0.1 Mpc} < 10^{-22}$G. Our previous estimate of the generated magnetic field consequently does not violate this bound, but the field strength is very low and may not suffice to seed large scale magnetic fields if not enhanced sufficiently. In \cite{Caprini09} it was found that an inverse cascade mechanism could transfer some field strength from small to larger scales thus partially escaping the effects of plasma damping. The inverse cascade mechanism requires a non-vanishing helicity of the primordial magnetic field, as one can expect in the presented scenario due to the large baryon asymmetry, thus one may still to successfully seed large scale magnetic fields fields at the QCD phase transition.

\section{gravitational waves}
The final signal of the QCD phase transition that we would like to discuss are gravitational waves. The process of nucleation and subsequent bubble collisions will stir hydrodynamic turbulence producing graviational waves in the process \cite{Witten84,KosowskyTurnerWatkins92,Kamionkowski94,Kahniashvili08,Huber08}. Again the Hubble parameter gives an important scale for the spectrum \cite{Witten84,Kahniashvili08}. Since the production mechanism is causal a peak frequency has to greater or equal to the Hubble frequency, $\nu_{peak}\geq\nu_H$. By how much this peak scale differs will depend on the details of the production mechanism and most importantly on the relevant time- and lengthscales that might  be significantly different from the Hubble frequency. Let us assume an exponential nucleation rate $\Gamma \propto \exp{(t/\tau)}$ with $\tau$ being the characteristic timescale for the nucleation process. Then the peak frequency of the spectrum due to the collision of bubbles will be given by
 \begin{equation}
\nu^B_{peak} \approx 4.0 \cdot 10^{-8} \mbox{Hz} \left(\frac{0.1 H^{-1}}{\tau}\right) \left(\frac{T^*}{150\mbox{MeV}}\right) \left(\frac{g^{eff}}{50}\right)^{1/6}
\end{equation}
where $T^*$ is the reheating temperature and the result is already redshifted to the present day frequency. It is common to denote a gravitational wave spectrum in terms of a characteristic strain amplitude which is definied in the following way 
\begin{equation}
	h_c(\nu)= 0.9 \cdot 10^{-18}\left(\frac{1\mbox{Hz}}{\nu}\right) \left(\frac{h_0}{0.7}\right) \left[\Omega_{gw}(\nu)\right]^{1/2}
\end{equation}
Using the results of \cite{Kahniashvili08} with the above estimates one arrives at a peak strain amplitude for the bubble collision peak of
\begin{equation}h_c(\nu^B_{peak})= 4.7 \cdot 10^{-15}
\left(\frac{\tau}{0.1 H^{-1}}\right)^2
\left(\frac{150\mbox{MeV}}{T^*}\right)
\left(\frac{50}{g^{eff}}\right)^{1/3}.
\end{equation}
Bubble collisions will also create hydrodynamic turbulence that will stir gravitational waves with a slightly lower peak frequency 
\begin{equation}
	\nu^T_{peak} \simeq 0.3\: \nu^B_{peak}
\end{equation}	
but with a higher peak amplitude 
\begin{equation}
	h_c(\nu^T_{peak})\simeq 2.1 h_c(\nu^B_{peak})
\end{equation}	
for a strongly first order phase transition \cite{Kahniashvili08}. The addition of magneto-hydrodynamic (MHD) turbulence could further boost the peak amplitude as shown in \cite{Caprini09A,Caprini10}. For frequencies lower than the Hubble frequency the spectrum should be uncorrelated white noise. The approximate shape of the strain amplitude
spectrum is then given by
\begin{eqnarray}
	h_c(\nu) \propto \nu^{1/2} ~~ &\mbox{for}& ~~ \nu < H \\
	h_c(\nu) \propto \nu^{n} ~~~~&\mbox{for}& ~~ \nu > \nu^B_{peak}
\end{eqnarray}	
where the spectral index $n$ should be at least -2 if the number of bubble collisions is low. If multi-bubble collisions play an important role the index is expected to be lower where simulations of the collision of vacuum bubbles as done in \cite{Huber08} find $n=-3/2$ while older simulations found $-2\leq n\leq-1$  \cite{KosowskyTurnerWatkins92,Kamionkowski94}. Note that a gravitational wave spectrum with $n=-1$ would be UV-divergent unless there is a cutoff \footnote{If $h(\nu)\propto\nu^{-1}$ then $\Omega_{gw}(\nu)$ is constant meaning that the total energy in the gravitational wave spectrum from the peak to the UV is given by $\Omega_{gw}=\int^\infty_{\nu_{peak}}\frac{dk}{k}\Omega_{gw}(k_{peak})$ and thus diverges logarithmically. We thank Ruth Durrer for bringing this point to our attention.}, so $n<-1$ for a realistic spectrum.

\begin{figure}[ht]
	\centering
		\includegraphics[width=0.48\textwidth]{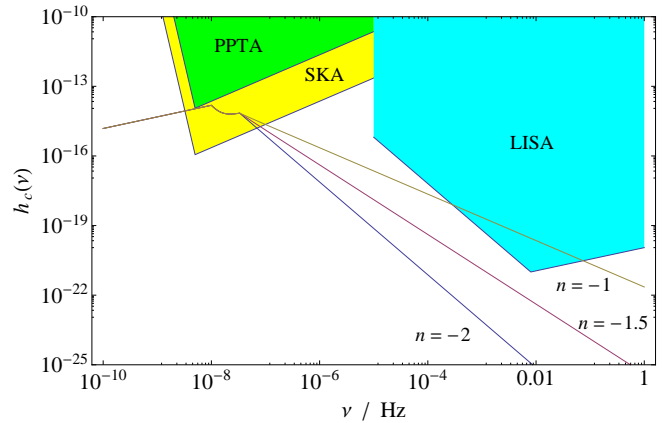}
		\caption{Largest strain amplitude spectrum at $\tau / H^{-1} = 0.12$ that is still compatible with the data of the Parks Pulsar Timing Array. A shorter duration of the phase transition reduces the amplidude and shifts the peak to higher frequencies. Detection with LISA would only be possible if multi-bubble collisions play a significant role.}
	\label{fig:gw}
\end{figure}

The Parks Pulsar Timing Array (PPTA) measures timing residuals in pulsar signals to put upper bounds on a stochastic gravitational wave background in a relatively narrow frequency band around $10^{-8}$ Hz \cite{Jenet06,Hobbs10}. The PPTA results already allow to limit the nucleation timescale with the presently available data to $\tau / H^{-1} < 0.12$. This limit will improve to $\tau / H^{-1} < 0.06$ for the full data of the Parkes Pulsar Timing Array project \cite{Jenet06} or even beyond that depending on the position of the peak \cite{Verbiest09,vanHaasteren11} since the sensitivity depends on the spectral index of the gravitational wave background. The results can be seen in figures \ref{fig:gw} where the expected spectrum of gravitational waves for the former case is shown with three different high frequency slopes and the approximate sensitivity regions of existing and future detectors. In \cite{Schettler10} the most optimistic results from \cite{Caprini10,Caprini09A} are compared with the most optimistic results from \cite{KosowskyTurnerWatkins92,Kamionkowski94} where the former have a slightly higher peak amplitude due to MHD turbulence while for the latter the contribution from multi-bubble collisions is stronger in the high frequency regime.\\

\begin{figure}[ht]
	\centering
		\includegraphics[width=0.48\textwidth]{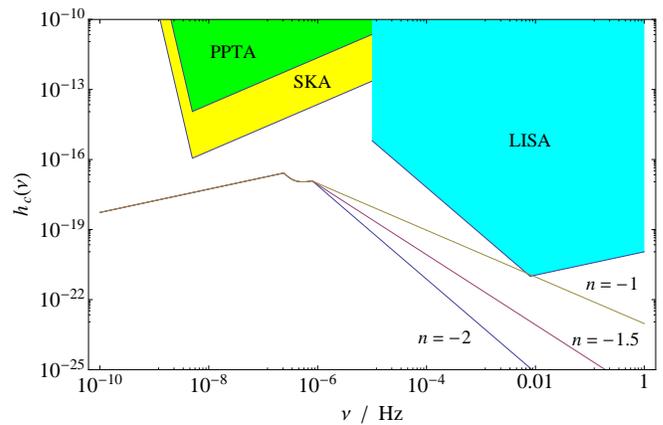}
		\caption{Strain amplitude spectrum for $\tau / H^{-1} = 0.005$ below which the signal would be unobservable with either SKA or LISA.}
	\label{fig:gw2}
\end{figure}

\noindent The planned Square Kilometer Array (SKA) will improve the sensitivity in $\Omega_{gw}(\nu)$ by about four orders of magnitude \cite{Kramer04}. Thus SKA will lower the bound on $\tau / H^{-1}$ by about an order of magnitude as visible in figure \ref{fig:gw}. If multi-bubble collisions are important detection via the spaceborne Laser Interferometer Space Antenna (LISA) could also be possible if the high frequency spectral index
$n\gtrsim -1.4$ and $\tau / H^{-1}\gtrsim10^{-2}$. In figure \ref{fig:gw2} the limiting case of a strain amplitude spectrum for $\tau / H^{-1} = 0.005$ is shown below which the signal would be unobservable with either SKA or LISA.\\
 
\noindent Furthermore it has been found that the QCD-phase transition will also leave a steplike imprint on the spectrum of primordial gravitional waves due to the strong reduction of the radiation degrees of freedom \cite{Schwarz98}. In \cite{Schettler10} this result was confirmed also for several lattice equations of state. Furthermore the effect of a little inflationary phase on the primordial spectrum was examined and a strong power-law suppression for frequencies larger than the Hubble frequency at the QCD phase transition was found.

\section{summary}
We have reexamined the idea of a little inflation at the QCD phase transition and extended the discussion as compared to our previous publication \cite{Boeckel10}. We found interesting cosmological implications such as the suppression of primordial density fluctuations up to dark matter mass scales of $M_{max} \sim 10^6 M_{\astrosun}$ relative to the large scale spectrum due to the change of the global equation of state. This could have interesting consequences for the physics of globular clusters and the emergence of the first stars and could also have an impact on the cuspy core density distribution of dark matter in small galaxies and the too large number of halo structures seen in standard structure formation. We found that the baryon density can actually be so large that one may even expect color superconducting phases to be present before the onset of the little inflation, this might pose an alternative route of investigation for the scenario.\\
We also discussed the production of primordial magnetic fields that may be strong enough to seed the presently observed galactic and extragalactic magnetic fields. Furthermore we addressed the production of a spectrum of gravitational waves around a peak frequency of $4 \cdot 10^{-8}$ Hz that may be observable via pulsar timing in the future \cite{Jenet06,Kramer04}.
Dark matter properties are also strongly affected as the annihilation
cross section for cold dark matter has to be up to nine orders of
magnitude lower to give the right amount of dark matter today, which
can be probed at the LHC by detecting the neutralino with an
unexpected low annihilation cross section.
The conditions in such a cosmological phase transition would then be closer to the situation in heavy ion collisions or even the centre of neutron stars than to the standard QCD phase transition
in the hot big bang scenario. Hence, the upcoming FAIR facility
would actually for the little inflation scenario be a probe for the physics of the early
universe.\\

\noindent We thank Rob Pisarski, Eduardo Fraga, Ruth Durrer, Chiara Caprini and Arthur Kosowsky for useful comments and discussions.\\

\noindent This work is supported by the Bundesministerium f\"ur Bildung und Forschung (BMBF) under grant FKZ 06HD9127, by the German Research Foundation (DFG) within the framework of the excellence initiative through the Heidelberg Graduate School of
Fundamental Physics (HGSFP) and through the Graduate Program for
Hadron and Ion Research (GP-HIR) by the Gesellschaft f\"ur
Schwerionenforschung (GSI), Darmstadt.
\section{appendix}
\label{appendix}
\noindent Now let us discuss the analytic solutions to the system of differential equations for the perturbations
\begin{eqnarray}
\delta_i^\prime & = & - \frac{3(c_{si}^2-w_i)}{a}\delta_i + \frac{k}{\mathcal{H}a} \hat\psi_i - 3 (1+w_i) \frac{\alpha}{a}\\
\hat\psi_i^\prime & = & - \frac{1-3w_i}{a}\hat\psi_i - c_{si}^2 \frac{k}{\mathcal{H}a} \delta_i - (1+w_i) \frac{k}{\mathcal{H}a}\alpha\\
\alpha & = & - \frac{\frac{3}{2}\left(1+ 3 c_{s}^2\right)}{\left(\frac{k}{\mathcal{H}}\right)^2+ \frac{9}{2}\left(1+w\right)}~\delta
\end{eqnarray}
for the most relevant cases. We will limit the discussion to stating the approximate equations of motion for the radiation and the dark matter component by directly giving the dominant analytic solutions.
\begin{table}[h]
\centering
\begin{tabular}{|l|l|}
\hline
\multicolumn{2}{|l|}{super-horizon: $k/\mathcal{H}\ll1$,}\\
\multicolumn{2}{|l|}{radiation domination: $c_s^2=w=1/3$,}\\
\multicolumn{2}{|l|}{$\alpha = - \delta_R/2$, $k/\mathcal{H} \propto a$}\\
\hline \hline
$\delta_R^\prime \simeq \frac{k}{\mathcal{H}a} \hat\psi_R + \frac{2}{a}\delta_R$ & $\delta_{DM}^\prime \simeq\frac{3}{2a}\delta_{R}$ \\
$\hat\psi_R^\prime  \simeq  \frac{1}{3} \frac{k}{\mathcal{H}a} \delta_R$ &$\hat\psi_{DM}^\prime \simeq - \frac{1}{a}\hat\psi_{DM} + \frac{1}{2}\frac{k}{\mathcal{H}a}\delta_{R}$\\
\hline
  $\delta_R = A \left(\frac{k}{\mathcal{H}}\right)^2$ &$ \delta_{DM} = \frac{3}{4} \delta_R$ \\
   $\hat\psi_R  =  \frac{A}{9}  \left(\frac{k}{\mathcal{H}}\right)^3$& $\hat\psi_{DM} = \frac{9}{8}\hat\psi_{R}$ \\
  \hline
\end{tabular}
  \caption{Equations of motion and dominant perturbation solutions for a radiation dominated universe in the super-horizon limit. The constant $A$ fixes the amplitude at horizon crossing and is scale independent for a Harrison-Zeldovic spectrum \cite{Schmid97,Schmid99}.} 
\label{suphubtable}  
\end{table}
\noindent First let us look at the growing super-horizon solutions in the case of a radiation dominated universe, which sets the initial condition for the numerical calculations. These solutions set the relevant initial conditions for our numerical calculations. The approximate equations of motion and the resulting solutions are summarized in table \ref{suphubtable}.\\

\noindent \textbf{Inflationary solutions}\\
\noindent Now we want to additionally find the solutions for the inflationary phase in the different spectral regimes. For the inflationary regime it will be most important to examine the case of $q=3$ because radiation will be less abundant than matter soon after the onset of inflation for relevant inflation lengths. First of all we need the mean quantities
\begin{equation}
1+ w \simeq \left(\frac{a_i}{a}\right)^3, ~~ \delta \simeq \delta_{DM} \left(\frac{a_i}{a}\right)^3, ~~ c_s^2 \simeq \frac{1}{3}, ~~ w \simeq -1\label{meanquantities}
\end{equation}
and we also need to remember that $k/\mathcal{H} \propto1/a$ in the following. Now let us examine the two most relevant spectral regimes namely the intermediate and the unaffected regime. For the spectral range we have examined numerically none of the modes will stay sub-Hubble sufficiently long during inflation to approach an analytic limit. This would only be the case for modes that stay similar or even below the Hubble frequency for the whole duration of inflation. It turns out that the solutions in this case are combinations of Bessel functions that cannot be found by simple analytic means. Thus we skip a lengthy discussion for these modes and directly jump to the other two regimes that are relevant to the discussion and have analytic solutions that can be derived rather quickly.\\

\noindent First let us discuss the intermediate modes with $1\gg\frac{k}{\mathcal{H}}\gg(1+w)^{1/2}$.
\begin{table}[ht] 
\centering
\begin{tabular}{|l|l|}
\hline
\multicolumn{2}{|l|}{intermediate modes: $1\gg\frac{k}{\mathcal{H}}\gg(1+w)^{1/2}$,}\\
\multicolumn{2}{|l|}{inflationary phase: $c_s^2\simeq1/3$, $w\simeq-1$, $k/\mathcal{H} \propto1/a$,}\\
\multicolumn{2}{|l|}{$\alpha \simeq - 3\left(\frac{\mathcal{H}}{k}\right)^2\left(\frac{a_i}{a}\right)^3 \delta_{DM}$}\\
\hline \hline
$\delta_{R}^{\prime} \simeq \frac{4}{3}\delta_{DM}^{\prime} $ & $\delta_{DM}^{\prime} \simeq 9\left(\frac{\mathcal{H}}{k}\right)^2\frac{a_i^3}{a^4} \delta_{DM}$ \\
$\hat\psi_{R}^\prime \simeq \frac{1}{3}\frac{k}{\mathcal{H}}\delta_{DM}^{\prime}$ &$\hat\psi_{DM}^\prime \simeq - \frac{1}{a}\hat\psi_{DM}$\\
\hline
  $\delta_{R} = \frac{4}{3} \delta_{DM}$ &$ \delta_{DM} = C_1 \exp{\left[-9\left(\frac{\mathcal{H}}{k}\right)^2\left(\frac{a_i}{a}\right)^3\right]}$ \\
   $\hat\psi_{R} = \left(\frac{k a}{3\mathcal{H}a_i}\right)^3 \delta_{DM}$& $\hat\psi_{DM}=\frac{C_2}{a}$ \\
  \hline
\end{tabular}
  \caption{Equations of motion and dominant perturbation solutions for intermediate modes during the short inflationary phase, $C_1$ and $C_2$ are constants.} 
\label{Intermediatetable}  
\end{table}
The approximate equations of motion in this regime and the corresponding solutions are summarized in table \ref{Intermediatetable}. The solutions for the dark matter perturbations imply that $\delta_{DM}$ and $\delta_R$ will be frozen very quickly until the end of inflation and approach a constant value. Note that $(1+w)^{1/2}$ drops quicker than $\frac{k}{\mathcal{H}}$ so any mode that enters this regime stays there until the end of inflation.\\

\noindent Now let us turn to the unaffected modes that are given by the condition $1\gg(1+w)^{1/2}\gg\frac{k}{\mathcal{H}}$. The corresponding approximate equations of motion and the resulting solutions in this regime are again summarized in table \ref{Unaffectedtable}.
\begin{table}[h] 
\centering
\begin{tabular}{|l|l|}
\hline
\multicolumn{2}{|l|}{unaffected modes: $1\gg(1+w)^{1/2}\gg\frac{k}{\mathcal{H}}$,}\\
\multicolumn{2}{|l|}{inflationary phase: $c_s^2\simeq1/3$, $w\simeq-1$, $k/\mathcal{H} \propto1/a$,}\\
\multicolumn{2}{|l|}{$\alpha \simeq- \frac{2}{3}\delta_{DM}$}\\
\hline \hline
$\delta_{R}^{\prime} \simeq \frac{4}{3}\delta_{DM}^{\prime} $ & $\delta_{DM}^{\prime} \simeq \frac{2}{a}\delta_{DM}$ \\
$\hat\psi_{R}^\prime \simeq - \frac{1}{3}\frac{k}{\mathcal{H}a}\delta_R+\frac{8}{9}\frac{k}{\mathcal{H}a}$ &$\hat\psi_{DM}^\prime \simeq - \frac{1}{a}\hat\psi_{DM} + \frac{2}{3}\frac{k}{\mathcal{H}a}\delta_{DM}$\\
\hline
  $\delta_{R} = \frac{4}{3} \delta_{DM}$ &$ \delta_{DM} = B a^2 \propto \left(\frac{\mathcal{H}}{k}\right)^2$ \\
   $\hat\psi_{R} =  \frac{4}{3} \hat\psi_{DM}$& $\hat\psi_{DM}=\frac{1}{3}\frac{k}{\mathcal{H}} B a^2\propto \left(\frac{\mathcal{H}}{k}\right)$ \\
  \hline
\end{tabular}
  \caption{Equations of motion and dominant perturbation solutions for unaffected modes during the short inflationary phase, $B$ is a constant.} 
\label{Unaffectedtable}
\end{table}
\noindent Note that the solutions for $\delta_R$ and $\delta_{DM}$ are exactly the same solution as for the radiation dominated super-horizon case as found before. Comparing these results to the analytic super-horizon solutions in the radiation dominated universe we find that $\delta_{DM}$ and $\delta_{R}$ have the same growing mode $\propto a^2$, thus the naming of the spectral region as "unaffected" is justified. On the other hand $\hat\psi_{DM}$ and $\hat\psi_{R}$ grow only linearly with the scale parameter in contrast to a cubic growth in the radiation dominated super-Hubble case. This is actually necessary to keep the spectrum scale invariant on large scales \cite{Schmid99} since this requires $\delta_i/\hat\psi_i \propto k/\mathcal{H}$. The latter keeps the amplitude at horizon entry independent of the wavenumber for large scales.\\
Furthermore the comparison between the intermediate and the unaffected cases shows that one may expect a
relative suppression in the fluctuations below the unaffected part of the spectrum because the intermediate modes are frozen in while the unaffected modes grow in the respective limit.

\end{document}